# Contactless Series Resistance Imaging of Perovskite Solar Cells via Inhomogeneous Illumination

Arman Mahboubi Soufiani[1*], Yan Zhu[1], Nursultan Mussakhanuly[1], Jae Sung Yun[1], Thorsten Trupke[1], Ziv Hameiri[1]

[1]School of Photovoltaic and Renewable Energy Engineering, University of New South Wales, Sydney 2052, Australia

E-mail: a.mahboubisoufiani@unsw.edu.au

## Abstract

A contactless effective series resistance imaging method for large area perovskite solar cells that is based on photoluminescence imaging with non-uniform illumination is introduced and demonstrated experimentally. The proposed technique is applicable to partially and fully processed perovskite solar cells if laterally conductive layers are present. The capability of the proposed contactless method to detect features with high effective series resistance is validated by comparison with various contacted mode luminescence imaging techniques. The method can reliably provide information regarding the severeness of the detected series resistance through photo-excitation pattern manipulation. Application of the method to sub-cells in monolithic tandem devices, without the need for electrical contacting the terminals, appears feasible.

# Introduction

The remarkable developments in the fabrication of high-efficiency single-junction perovskite solar cells (PSC) with power conversion efficiencies greater than 25% [1], [2] have generated significant interest in the commercialisation of this promising technology. Nevertheless, the commercialisation of PSC still faces key challenges that need to be addressed to make this technology competitive: (1) long-term stability [3], [4], and (2) device area upscaling [5], [6]. Critical aspects of PSCs upscaling are obtaining homogeneous perovskite absorber layer properties and selective contacts [7], [8], as well as minimal resistive losses [5], [9]. Development of suitable characterisation techniques, which can reliably and rapidly identify the factors contributing to these two main issues, is essential [10]–[12].

Camera-based electroluminescence (EL) [13] and photoluminescence (PL) [14] imaging (ELi and PLi) were introduced to photovoltaic (PV) research and development (R&D) more than a decade ago and are now established as main characterisation techniques in both research laboratories [15] and in high volume production-lines of silicon solar cell. They are used across the entire value chain including silicon bricks [16], wafers [14], cells [17] and modules [18], [19], significantly contributing to process optimisation of silicon PV devices. A wide range of electrical parameters have been quantified and spatially resolved, including: implied open-circuit voltage (i$V_{OC}$) [20], ideality factor [21], dark saturation current-density ($J_0$) [22], [23], shunt resistance [24]–[26], series resistance ($R_s$) [22], [27]–[30], and others [13], [31].

The first adaptions of luminescence-based imaging to PSCs were in 2015 [32], [33]. Since then, a broad range of applications has been developed. ELi and PLi have been used to study charge-carrier recombination and resistive losses [34], to detect pinholes [35], to assess the quality and uniformity of the perovskite layer [12], [36], and more [37], [38]. They have been also used to image i$V_{OC}$ [39]–[43], spatial variation of the optical bandgap [44], current transportation [10], [42], and $R_s$ [45], [46].

Spatially resolved $R_s$ images are commonly extracted through two main approaches [47], which involve either ELi [25] or PLi with current extraction (PL$_{CE}$) via the device terminals [27], [30]. A common EL-based method is based on the collection of images at two voltage biases. One image is acquired at a low voltage bias (forward current) at which $R_s$ effects are negligible [30], allowing a reasonable estimate of the local diode parameter (i.e. $J_0$) [28], while the second image is obtained at a high voltage bias. An alternative iterative approach was later developed eliminating the need for the low-bias EL image, while achieving $R_s$ images with improved quality [48]. An elegant $R_s$ imaging method based on PL image acquisition with simultaneous current extraction was developed by Kampwerth *et al.* [27]. Both the EL and PL$_{CE}$ approaches provide *effective* $R_s$ values, which lump the contact and transport resistances for each cell location in the current path between the two terminals. Furthermore, and most relevant for the following discussion, both approaches are performed in a *contacted* mode, requiring a completed metallised device.

Contactless characterisation methods are generally advantageous, since they can be implemented in production with much higher measurement throughput compared to contacted methods. They also significantly reduce the risk of damaging the devices. Non-contact $R_s$ imaging was first developed and demonstrated for silicon solar cells by Kasemann *et al*. [49]. Without the necessity of electrical contacting of the device, valuable information about areas suffering from high effective $R_s$ can be gained. Noticeably, this technique allows the distinction of series resistance losses between the front and rear contacts and the absorber [50], as long as the carrier collecting layers have sufficient lateral conductivity. If not, metal or transparent conductive oxide electrodes are needed to facilitate lateral carrier transport. In the PSCs world, this contactless method can further avoid the consumption of expensive metal electrodes such as gold, as a completed device is not required.

Here, we present a contactless PLi method for the extraction of qualitative effective $R_s$ images for PSCs. A flexible spatially non-uniform photo-excitation pattern is applied to create alternating illuminated and non-illuminated areas across the cell. The resulting lateral balancing currents within the sample under test give rise to *contactless* PL images with current extraction ($PL_{CE}^{contactless}$) in the illuminated regions and *contactless* EL ($EL^{contactless}$) images in the non-illuminated regions, respectively. Subsequent processing of the obtained luminescence images provides qualitative information about the spatial distribution of the voltage drop across effective $R_s$. It is demonstrated that through this inhomogeneous-illumination-based contactless PLi approach, the severity of $R_s$ regions can be identified and studied. A key advantage of this contactless approach is the possibility to investigate $R_s$ at early fabrication stages.

## Background

The principle of the method implemented in this study follows that of Kasemann *et al*. [49]. In its original form, at least three $PL_{OC}$ images are collected under open-circuit condition [49]. Physical shading is used in Kasemann's method to create the non-illuminated regions and to stimulate the lateral balancing currents from illuminated to non-illuminated regions. In the modified method of Zhu *et al*. [51], non-uniform illumination is achieved using a digital micromirror device (DMD), which has the advantage that luminescence emission can be collected from the entire cell area, including the non-illuminated parts.

The detected luminescence intensity in a solar cell is related to its diode voltage, *V*, at any operating point, through [30], [52]:

$$\emptyset_{PL(EL)}(x,y) = C(x,y) \times e^{\left(\frac{V(x,y)}{V_{th}}\right)} \tag{1}$$

which is relevant for both PL and EL measurements. The detected position-dependent wavelength-integrated intensity is denoted as $\emptyset_{PL(EL)}(x,y)$. $V_{th}$ is the thermal voltage at the sample temperature. $C(x, y)$ is a calibration constant that depends on the configuration of the measurement setup and on the optical properties of the sample. Note that although $C(x, y)$ can have a voltage dependency, it is assumed to be injection-independent under the operating conditions used in this study.

Assuming a simple local one-diode model, the voltage difference between the local diode voltage $V(x,y)$ and the terminal voltage $V_T$ is related to the local series resistance $R_s(x,y)$ via:

$$V_T - V(x,y) = V_s(x,y) = R_s(x,y) \times J(x,y) \tag{2}$$

where $V_s(x,y)$ is the voltage drop associated with current flow across the effective series resistance associated with each specific cell location. $J(x,y)$ is the local current density that must be quantified for quantitative measurements of $R_s$.

By combining PL images taken with complementary illumination patterns, we can generate a PL image of the full cell that is equivalent to a PL image taken with current extraction via the terminals in a contactless mode ($PL_{CE}^{contactless}$ image). Dividing this combined image by a PL image taken with full area illumination ($PL_{OC}$ image), using Eq. (1), the diode voltage difference can be obtained:

$$\Delta V(x,y) = V_{th} \times \ln\left(\frac{\phi_{PL}^{CE}(x,y)}{\phi_{PL}^{OC}(x,y)}\right) \qquad (3)$$

The voltage difference $\Delta V(x,y)$ represents the difference in the diode voltage between two operating points at each cell location, the first operating point representing $V_{OC}$ conditions, with no current extraction and thus with rather uniform voltage distribution across the cell, while the second operating point is under illumination and simultaneous current extraction. In the second scenario, areas with increased $R_s$ remain at a higher local diode voltage since carrier extraction is less efficient. As a result, the value of $\Delta V(x,y)$ is larger in those areas.

If the two operating points discussed above are chosen in such a way that similar excess carrier density is present in the two measurements, then this division removes, or at least reduces, prominent features that are associated with defect-mediated recombination losses.

Higher contrast $R_s$ images are obtained if the luminescence emission can be also collected from the shaded (non-illuminated) regions [51]. The signal detected from the non-illuminated (illuminated) regions is similar to the luminescence distribution expected in EL images ($PL_{CE}$ images), as it originates from injected (extracted) current, flowing in from (out of) the illuminated cell areas (see **Figure 1**).

In this study, a DMD is employed to alternately create the illuminated/non-illuminated patterns. Taking the ratio of the combined luminescence images of the illuminated parts (the $PL_{CE}^{contactless}$ images; see **Figure 1**) to the combined luminescence images of the non-illuminated parts (the $EL^{contactless}$ images; see **Figure 1**) provides an image of the full cell containing information proportional to the qualitative $R_s$ images with enhanced contrast. This improvement is largely because $R_s$ features appear with inverted contrast in the two images (luminescence signal associated with the regions with high (low) effective $R_s$ in the $PL_{CE}$ image is amplified (reduced) when divided by the corresponding low (high) intensity signal of the related EL images), whereas recombination features appear with the same contrast (dark in both images) and therefore the latter are largely removed in the ratio image. The local diode voltage difference is calculated through:

$$\Delta V(x,y) = V_{th} \times \ln\left(\frac{\phi_{PL}^{CE}(x,y)}{\phi_{EL}(x,y)}\right) \qquad (4)$$

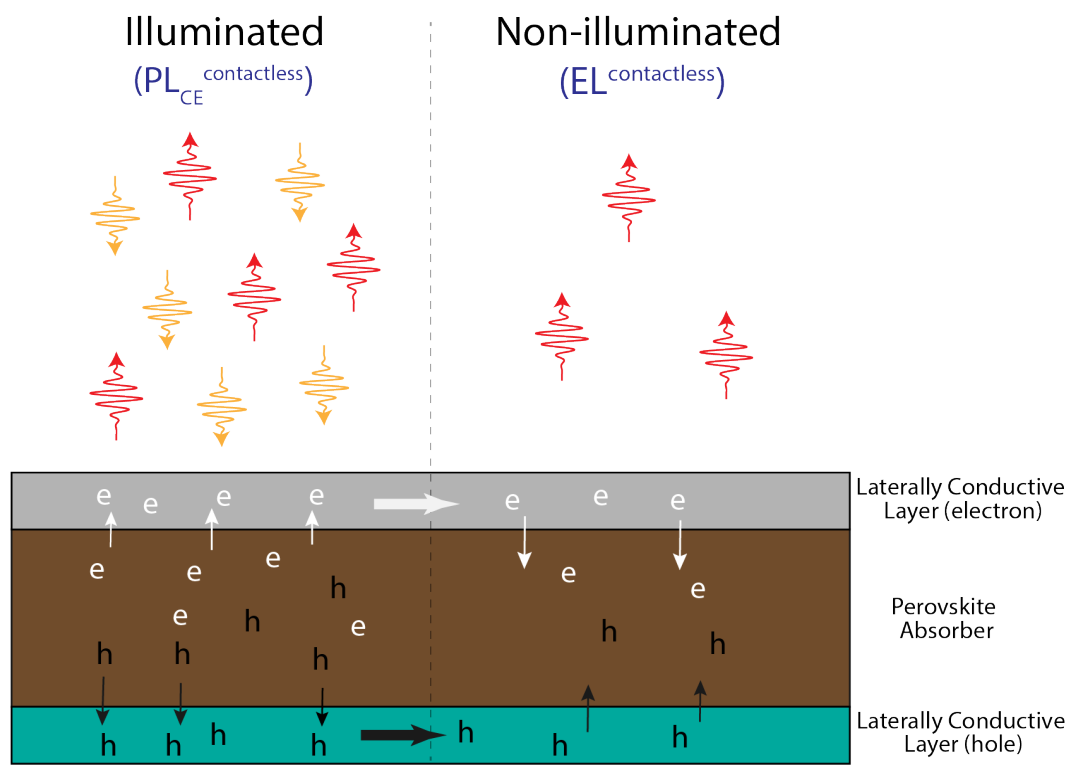

**Figure 1 Working principle.** Schematic illustrating the process through which the contactless PL$_{CE}$ and EL signals are created upon inhomogeneous illumination. The yellow and red wave packets represent excitation and luminescence, respectively.

Measurement Setup

**Figure 2A** illustrates the measurement setup. The inhomogeneous illumination source consists of an excitation light source, a DMD chip, a colour wheel, and lenses for flexible illumination pattern creation. Depending on the configuration of the micromirrors, the light is either reflected off the mirrors and then projected onto the sample or is directed toward a beam dump. The output spectrum of the excitation source, before passing through a 675 nm shortpass filter (given in the inset of **Figure 2B**), is incident on a 45° cold mirror. The mirror reflects >90% of the incident illumination with wavelengths shorter than 675 nm (see 'Reflect.' in the inset of **Figure 2B**) towards the sample. The luminescence spectrum emitted from the sample then passes through the same mirror (with >90% transmission for wavelengths longer than 725 nm; see 'Trans.' in the inset of **Figure 2B**) and a set of filters before being detected by the sCMOS (scientific complementary metal–oxide–semiconductor) camera. It is noted that the tail of the excitation spectrum at wavelengths longer than 725 nm, which can partly transmit through the longpass filters in front of the camera and being detected, is significantly attenuated beforehand and does not interfere with the sample's strong luminescence signal. Only a small fraction of this tail is first reflected off the cold mirror (cold mirror reflectance is at its minimum for >725 nm) of which less than 20% is reflected off the device and is then transmitted through the cold mirror. A typical luminescence spectrum from the PSCs investigated in this study, as well the transmission of the optical filter stack that is placed in front of the camera are presented in **Figure 2B**.

In this study we used PSCs with the following device structure: FTO/c-TiO$_2$/mp-TiO$_2$/Cs$_{0.05}$FA$_{0.80}$MA$_{0.15}$Pb(I$_{0.85}$Br$_{0.15}$)$_3$/Spiro-OMeTAD/Au. The fabrication procedure is provided in the Supplementary Information, as well as the measured illuminated current-voltage (*J-V*) curve (**Figure S1**).

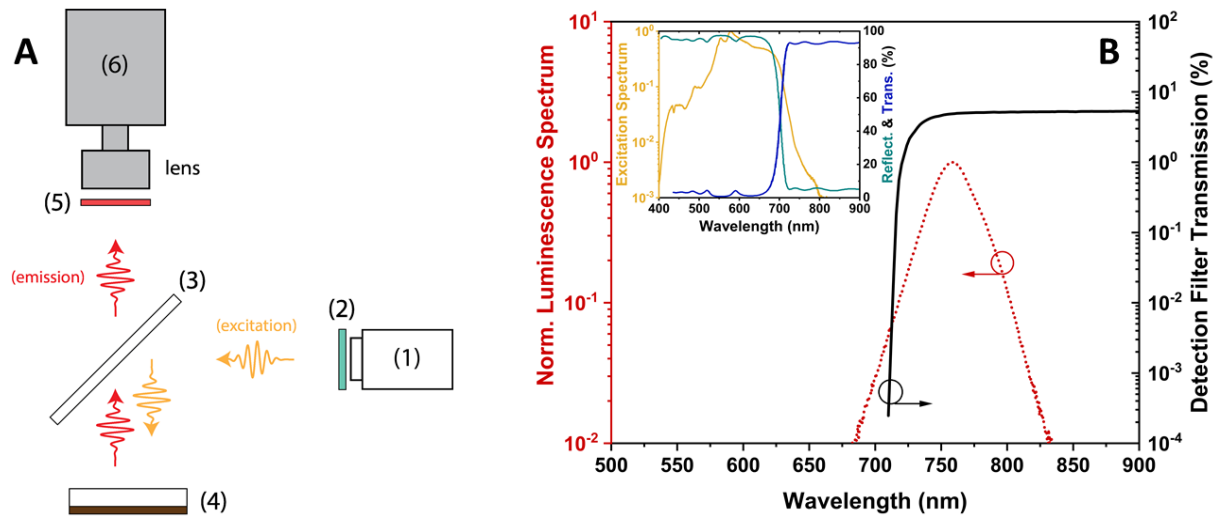

**Figure 2 Measurement setup.** A) Schematic of the measurement setup: (1) inhomogeneous illumination source consisting of the light source, DMD chip, colour wheel, and lenses, (2) shortpass filter, (3) 45° cold mirror, (4) PSC, (5) longpass and neutral density filters, and (6) sCMOS camera. B) Dotted line: spectral luminescence of a representative device; solid line: overall transmission spectrum of the optical filter set used in front of the camera lens. The inset includes the spectrum of the excitation source before passing through a shortpass filter before incident on the cold mirror, as well as the reflection and transmission spectra of the 45° cold mirror.

## Results and Discussion

It is noted that all luminescence images used in this study were collected immediately after the initiation of the external excitation of the solar cell, whether that being through photoexcitation or voltage biasing/current injection. This is mainly to avoid complications associated with the potential transient effects observed in PSCs, which can induce reversible and irreversible changes to the device performance [34], [53].

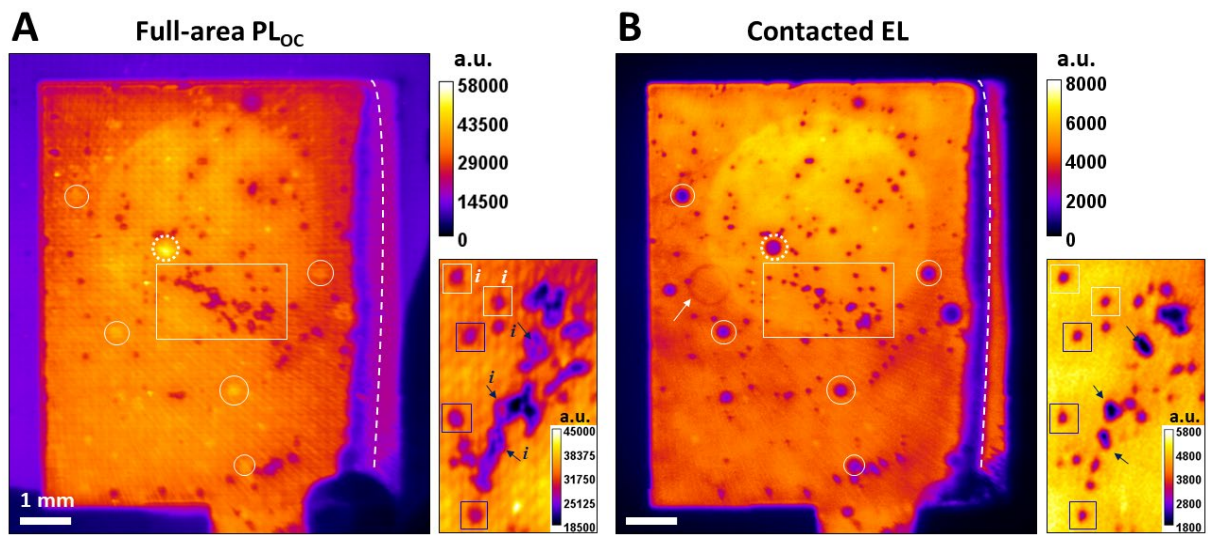

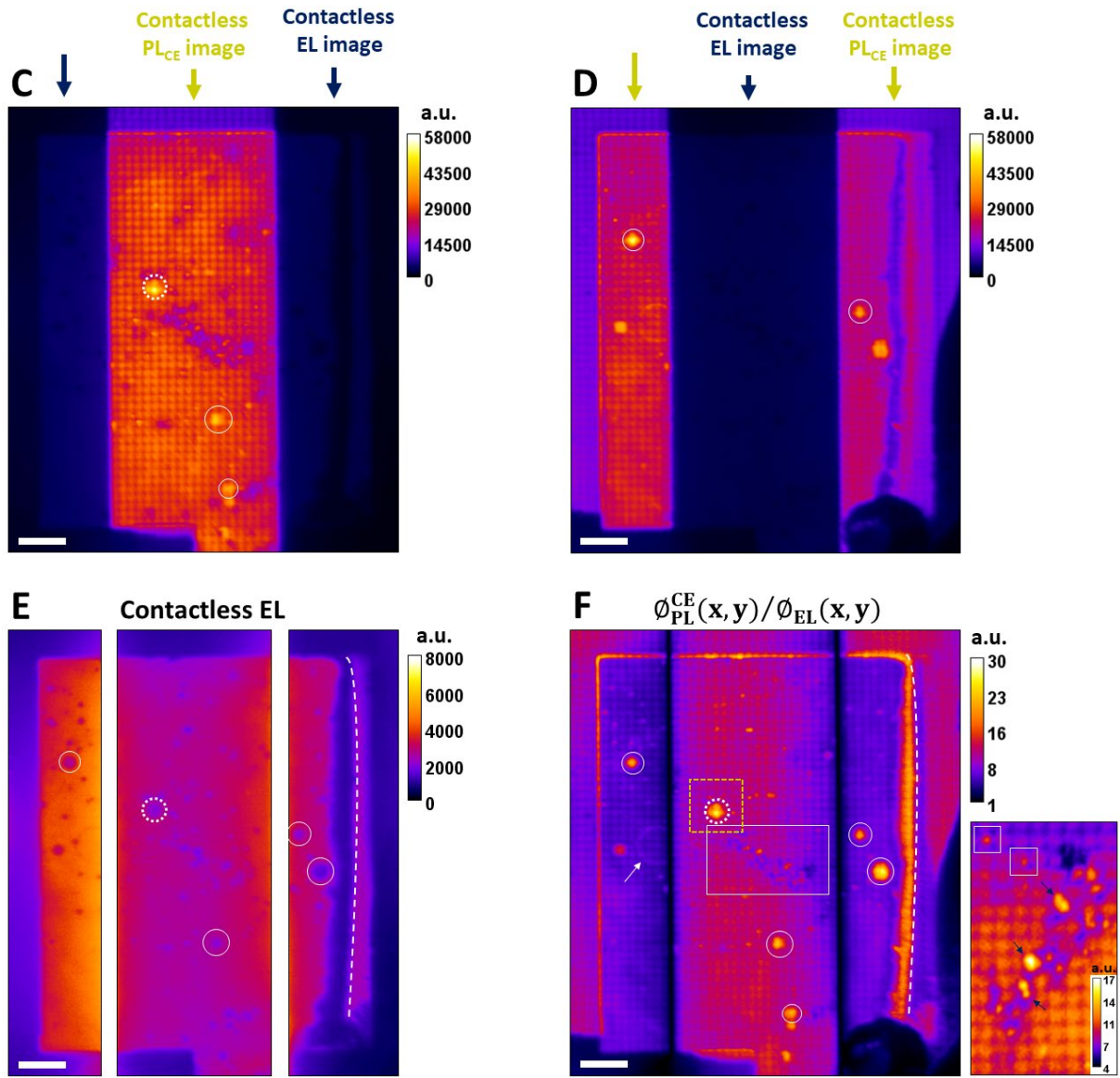

**Figure 3 Demonstration of contactless qualitative $R_s$ imaging on a 7 mm × 9 mm PSC.** A) Full-area $PL_{OC}$ image of the PSC illuminated at ~0.75 Sun equivalent photon flux (*J-V* curve is given in **Figure S1**). B) EL image of the same cell, measured at an applied bias of 1.15 V, close to the device's $V_{OC}$. C) and D) are PL images of the device, partly illuminated at ~0.75 Sun equivalent photon flux. Note that the same colour scale is used for (A), (C), and (D). E) Contactless EL image of the full device obtained by combining the non-illuminated parts of the images (C) and (D). F) Ratio image of the appended contactless $PL_{CE}$ images and the contactless EL images. The colour scale of the inset images in panels (A), (B), and (F) are different from the corresponding main images and are adjusted to improve visibility of the features. The scalebar shown at the bottom left of the images is 1 mm.

<u>Full area PL and EL imaging</u>

A $PL_{OC}$ image of the PSC at ~0.75 Sun is shown in **Figure 3A** (the observed meshed pattern is an artefact created by the excitation light source). Several circular features with lower emission intensities can be observed, possibly due to higher non-radiative recombination within the bulk perovskite and/or perovskite-contact interfaces, which could arise from pinholes or non-uniform coating. These are exemplified with squares in the inset of **Figure 3A**. The latter represents the region of interest (ROI) from the centre of the $PL_{OC}$ image specified with a white rectangle.

The EL image of the same PSC taken at a bias voltage of 1.15 V is presented in **Figure 3B**. As the used bias voltage is close to $V_{OC}$, the image is expected to be strongly influenced by $R_s$ [32], [34]. The presence of similar low-intensity features (marked with squares) in the corresponding EL image further supports their association with highly recombination active sites, at least at their peripheries (this will be further discussed in the next section).

However, within the same rectangular ROI there are island-like areas in the $PL_{OC}$ image, indicated with arrows (and with black '*i*' in the inset images) that show high PL intensity in the middle of the features with low intensities in their surroundings, yet they appear as totally low-intensity features in the relevant EL image. As will be discussed below, the middle regions of these features seem to be isolated regions with high effective $R_s$ while the surrounding of the islands are sites with high non-radiative recombination.

Distinctive areas in the EL image (marked with white circles) with lower luminescence intensities with respect to their surroundings can be identified. Interestingly, the same regions appear with high intensity relative to their surroundings in the corresponding $PL_{OC}$ image of **Figure 3A**. The cause of this anti-correlation behaviour is not well understood and requires further investigation. This could be due to a poor local physical contact between the absorber and the selective contact (hence high $R_s$), as likely free-standing perovskite surfaces are not impacted by high surface recombination associated with direct contact with spiro-OMeTAD or $TiO_2$ [34], [54].

Another low-intensity ring-like feature in the EL image, indicated with an arrow on the left-hand side of the image, which is absent in the $PL_{OC}$ image, is also a possible region with high resistive losses.

The most evident feature in both the $PL_{OC}$ and EL images of this device is the low intensity wavy stripe on the right-hand side of **Figures 3A** and **3B**. A dashed curved line is added where the edge of this feature begins in the EL image as a guide to the eye. The specified edge of this dark feature in the EL image overlays with a relatively bright area of the $PL_{OC}$ image, a little away from the right of the edge of the severely low-intensity region of the $PL_{OC}$ image. Thus, we speculate that at least part of this long feature appearing in the EL image is affected by high $R_s$, whereas parts of it towards its left-hand side could be suffering from high non-radiative recombination losses consistent with the $PL_{OC}$ results.

The relatively large circular area that appears bright in both the $PL_{OC}$ and EL images, respectively, corresponds to the area of the aperture used for the *J-V* measurements, indicating an increased charge-carrier density and, thus, higher lifetime in this region after the *J-V* scans. This shows a strong sensitivity of PSC devices to light-induced changes, as reported previously [34].

### PL imaging with inhomogeneous illumination

Next, we applied the inhomogeneously-illuminated PLi approach to evaluate the high-effective $R_s$ regions identified above. The illumination pattern – created by the DMD – divides the PSC area into three sections, alternately illuminated and non-illuminated. The total illuminated area approximately equates to the non-illuminated area. It is noteworthy that the ability to generate arbitrary illumination patterns can be readily exploited with this technique to vary the extent of current extracted from selected areas of the device. **Figure 3C** shows the PL image of the device with only the middle part (approximately half of the area) of the PSC being illuminated while the other two sides, each approximately a quarter of the cell area, are non-illuminated. **Figure 3D** presents the image measured with the inverse illumination pattern. Under these conditions at most half of the generated excess charge-carriers can be extracted from the illuminated section via lateral balancing currents.

The contactless EL image of the full device (generated by combining the two images acquired with partial illumination) is depicted in **Figure 3E**, in which the non-illuminated parts of the device in **Figures 3C** and **3D** were appended. The features in the EL$^{contactless}$ image correspond well with those in the contacted EL image of **Figure 3B**. For instance, note the circular dark features that are highlighted in the images with white circles. The higher EL intensity collected from the left part of **Figure 3E** as compared to the right part is an artefact that is likely due to the presence of the relatively large wavy stripe feature acting as a sink for the injected current.

Using **Figures 3C** and **3D**, both a contactless EL and a contactless PL image with current extraction were generated, from which a qualitative effective $R_s$ image was calculated. **Figure 3F** shows the ratio between the combined full area $PL_{CE}^{contactless}$ image and the combined full area EL$^{contactless}$ image. Similar to $PL_{CE}$ images, regions with high $R_s$ appear brighter in this ratio image. As can be seen, the proposed contactless PL-based method detects all the regions that were identified as regions suffering from high effective $R_s$. These include the circular spots specified with white circles, the island-like features within the rectangular ROI (and in other regions of the device), and the wavy stripe feature along the right-hand side of the device. The proposed approach also identifies the ring-like feature as a $R_s$-affected region, as suspected earlier from the comparison between **Figures 3A** and **3B**.

The high contrast associated with the low charge-carrier lifetime in the surroundings of the island-like features in the rectangular ROI (see insets of **Figures 3A** and **3B**) is reduced in the ratio of the contactless luminescence images (**Figure 3F**). This indicates that by implementing this approach, if the contactless images are not operating at distinctively different injection levels, the non-uniformity associated with charge-carrier recombination within the bulk perovskite and/or at the perovskite-selective contact interfaces can be removed or at least significantly reduced and mainly $R_s$ features are presented.

Furthermore, this approach revealed some of the features, such as those exemplified with white squares in the rectangular ROI, appearing with low luminescence signal in both $PL_{OC}$ and contacted EL images (insets of **Figures 3A** and **3B**), as sites with high effective $R_s$. These seem to behave like the island-like features, although could not be resolved from the $PL_{OC}$ image (all such features are indicated with white '*i*' in the inset of **Figure 3A**). A possible explanation is the association of these features with pinholes, present for instance in the spin-coated Spiro-OMeTAD contact layer.

Contactless PL imaging with increased current extraction

Although the proposed contactless method does not provide quantitative information about $R_s$, it allows identification of the severity of the $R_s$ features. This capability is demonstrated in this section. The inhomogeneous illumination pattern was modified such that only a small region, marked with a dashed square in **Figure 3F** (approximately 1 mm × 1 mm), is illuminated while the rest of the device is non-illuminated (see **Figure S2**), forcing this illuminated region to operate under high current extraction condition. The panels of **Figure 4** display the cropped luminescence images of this region: (A) $PL_{CE}^{contactless}$ image, (B) $EL^{contactless}$ image, (C) $\Delta V_s$ (natural logarithm of the ratio image between panels **4A** and **4B**), (D) $\Delta V_s$ image of the same region cropped from **Figure 3F**, (E) $PL_{OC}$ image of the uniformly illuminated full device, and (F) contacted EL image at 1.15 V.

The area marked with a dashed white circle in **Figure 4C** shows no sign of the selected circular $R_s$ feature, whereas the same area shows high $R_s$ in **Figure 4D**, which was measured at a maximum of 50% of current extraction, as discussed above. The island-like regions with high resistive losses in this selectively illuminated region remain as bright spots in both effective $R_s$ images, indicating that these are potentially fully isolated regions with high absolute $R_s$ compared to the circular feature with lower absolute $R_s$. These results highlight the potential of the proposed contactless PL-based imaging approach in distinguishing regions with different absolute $R_s$.

We reiterate the point that by taking the ratio of the contactless images, the resulting equivalent contactless $R_s$ image removes to a large extent the contribution associated with charge-carrier recombination. The high recombination region, with low PL signal, surrounding the island-like feature, indicated with '$i$' in the $PL_{OC}$ image of **Figure 4E**, is eliminated in the $R_s$ image of **Figure 4C**.

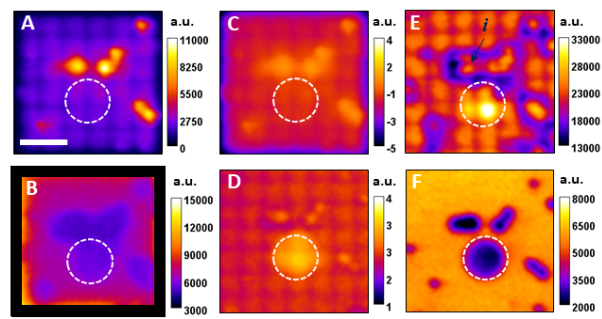

**Figure 4 Contactless qualitative $R_s$ imaging with increased current extraction.** (A) Luminescence image of the same device where a square-like area, highlighted by a dashed square in Figure 3F, is selectively locally illuminated while the rest of the device is non-illuminated, providing a $PL_{CE}^{contactless}$ image of the selectively illuminated area. (B) Image acquired with the inverse illumination pattern, providing $EL^{contactless}$. (C) is the image proportional to $\Delta V_s$. (D) $\Delta V_s$ image of the same region obtained from Figure 3F. (E) and (F) are the PL image measured with full area illumination and the contacted EL image, respectively. The scalebar shown at the bottom left of image A is 400 μm.

Contacted PL imaging with current extraction

To verify the reliability of the contactless method, $PL_{CE}$ measurements were also performed in a *contacted* mode. **Figure 5A** presents a $PL_{CE}$ image acquired with ~0.75 Sun equivalent photon flux, in which current extraction is performed at a voltage about 150 mV higher than $V_{MPP}$. All the device regions impacted by increased effective $R_s$ highlighted in the contactless image of **Figure 3F** are consistently observed in **Figure 5A**. To evaluate the severity of the $R_s$ features discussed in the previous section through contactless approach, the extent of current extraction in the contacted mode was varied. As the operating point of the cell approaches short-circuit conditions, moving from above $V_{MPP}$ (**Figure 5A**) to ~300 mV below $V_{MPP}$ (**Figure 5C**), the circular $R_s$-affected regions (marked with white circles) and the ring-like features seen in **Figure 5A** almost disappear. Note the disappearance of the feature indicated with a dashed white circle, as the current extraction is increased, consistent with the results of **Figure 4**.

The contacted mode $PL_{CE}$ imaging results support the above interpretation of the variations in image features around $R_s$ regions observed in the images obtained with partial illumination: features that disappear as the extent of current extraction increases have lower absolute $R_s$ values compared to those that remain visible. The residual bright spots retained in **Figure 5C**, representing a PL image acquired with the cell operating close to short-circuit conditions, are likely fully isolated regions with high absolute $R_s$. It is noteworthy that it is also possible that such regions have a current extraction level dependent $R_s$ [46].

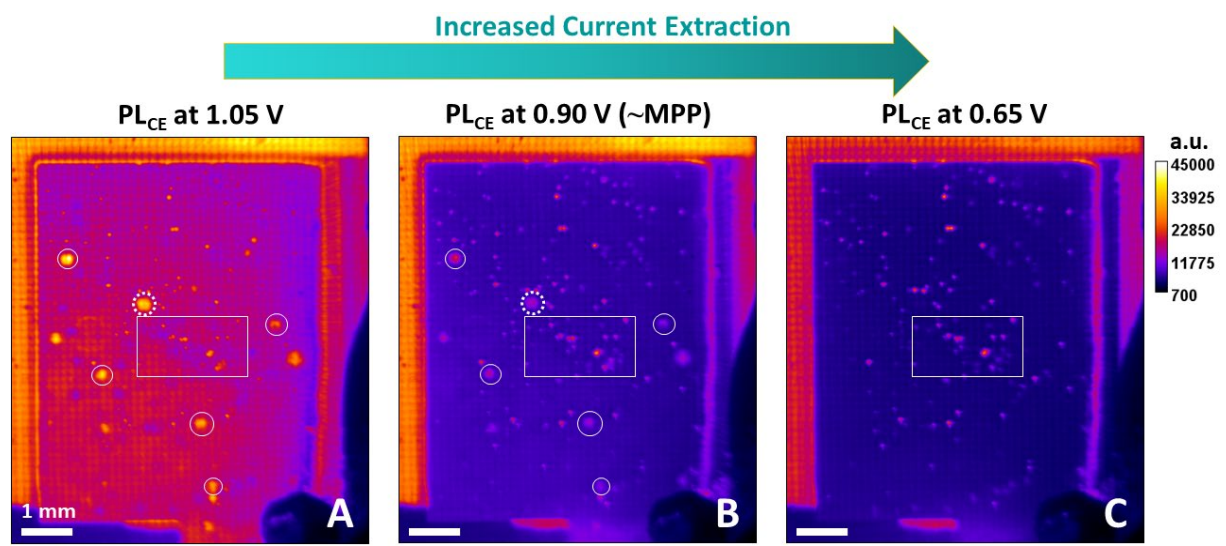

**Figure 5 Contacted mode $PL_{CE}$.** PL images with different current extraction levels of the same device measured in a contacted mode, $PL_{CE}^{contacted}$, with terminal voltages of (A) 1.05 V, (B) 0.9 V, and (C) 0.65 V. The PSC is illuminated at ~0.75 Sun equivalent photon flux. The $PL_{CE}^{contacted}$ images are in the same colour scale. The scalebar shown at the bottom left of the images is 1 mm.

## Conclusions

In summary, we demonstrated the application of PL-based contactless $R_s$ imaging on perovskite-based PV devices using inhomogeneous illumination. It was shown that the operating point of particular cell regions, in particular the fraction of photocurrent that is extracted during PL image acquisition, can conveniently be manipulated experimentally, via the sizes of the illuminated and non-illuminated areas, respectively. The resulting ability of this contactless technique to identify and differentiate $R_s$-affected regions with high and low absolute $R_s$ was confirmed qualitatively by comparison with photoluminescence images that were acquired with simultaneous current extraction in contacted mode. The convenient and fast access to regions with high effective $R_s$ provided by this contactless approach can be used for process optimization, in particular during the upscaling of PSCs to larger device areas. A potential future application of this method is its adaptation to fully contactless characterisation of sub-cells in monolithic tandem devices, where two critical PV parameters, $R_s$ and $iV_{OC}$ can be mapped in a non-destructive PL-based contactless mode.

## Acknowledgement

This work has been supported by the Australian Government through the Australian Renewable Energy Agency (ARENA) and the Australian Centre of Advanced Photovoltaics (ACAP). The views expressed herein are not necessarily the views of the Australian Government, and the Australian Government does not accept responsibility for any information or advice contained herein. Arman Mahboubi Soufiani acknowledges the funding support from ACAP (RG193402-I). Yan Zhu acknowledges the funding from ACAP (RG200768-G).